\documentclass[runningheads]{llncs}
\usepackage[T1]{fontenc}
\usepackage{graphicx}
\usepackage{hyperref}
\usepackage{color}

\usepackage{amsmath,amssymb}
\usepackage[capitalize,noabbrev]{cleveref}
\usepackage{subcaption} 
\usepackage{xcolor}

\usepackage{thmtools} 
\usepackage{thm-restate}

\usepackage{xspace}
\usepackage[most]{tcolorbox} 
\newtcolorbox{importantbox}
{ colback=gray!15, 
 colframe=black, 
 boxrule=0.8pt, 
 arc=2mm 
}

\newcommand{\ER}{\ensuremath{\exists \mathbb{R}}\xspace}
\newcommand{\R}{\ensuremath{\mathbb{R}}\xspace}
\newcommand{\A}{\ensuremath{\mathcal{A}}\xspace}
\newcommand{\C}{\ensuremath{A}\xspace}
\newcommand{\U}{\ensuremath{\mathcal{U}}\xspace}
\renewcommand{\L}{\ensuremath{\mathcal{L}}\xspace}
\newcommand{\NP}{\textsc{NP}\xspace}
\newcommand{\PSPACE}{\textsc{PSPACE}\xspace}

\newcommand{\CoinDraw}{\textsc{Penny Arrangement}\xspace}
\newcommand{\stretchability}{\textsc{Stretchability}\xspace}

\newcommand{\coinArrangement}{unit circle arrangement\xspace}
\newcommand{\intersectionsA}{inner intersections\xspace}
\newcommand{\intersectionsB}{outer intersections\xspace}
\newcommand{\bbox}{\ensuremath{B}\xspace}

\usepackage{environ}
\usepackage{tabularx}
\makeatletter
\newcommand{\problemtitle}[1]{\gdef\@problemtitle{#1}}%
\newcommand{\probleminput}[1]{\gdef\@probleminput{#1}}%
\newcommand{\problemquestion}[1]{\gdef\@problemquestion{#1}}%
\newcommand{\problemoutput}[1]{\gdef\@problemoutput{#1}}%
\NewEnviron{problemd}{
  \problemtitle{}\probleminput{}\problemquestion{}%
  \BODY%
  \par\addvspace{.5\baselineskip}
  \noindent
  \begin{tabularx}{\textwidth}{@{\hspace{\parindent}} l X c}
    \multicolumn{2}{@{\hspace{\parindent}}l}{\@problemtitle} \\
    \textbf{Input:} & \@probleminput \\
    \textbf{Question:} & \@problemquestion%
  \end{tabularx}
  \par\addvspace{.5\baselineskip}
}

\newcommand{\psline}[1]{\ensuremath{\ell_{#1}}}
\newcommand{\stline}[1]{\ensuremath{b_{#1}}}
\newcommand{\pscircUp}[1]{\ensuremath{\overline{a_{#1}}}}
\newcommand{\pscircDown}[1]{\ensuremath{\underline{a_{#1}}}}
\newcommand{\stcirc}[1]{\ensuremath{c_{#1}}}
\newcommand{\stcircUp}[1]{\ensuremath{\overline{c_{#1}}}}
\newcommand{\stcircDown}[1]{\ensuremath{\underline{c_{#1}}}}
\newcommand{\stcircUD}[1]{\ensuremath{\overline{\underline{c_{#1}}}}}
\newcommand{\pscircUD}[1]{\ensuremath{\overline{\underline{a_{#1}}}}}
\newcommand{\circsegUD}[1]{\ensuremath{\overline{\underline{x_{#1}}}}}

\begin{document}
\title{What can you draw with one coin?\\
\ER-completeness of Penny-Arrangements.
}
\titlerunning{What can you draw with one coin?}
\author{Lukas Bachmann\inst{1} \and
Lucas Meijer\inst{2} \and
Tillmann Miltzow\inst{2} \and Rosna Paul\inst{3}\and Soeren Terziadis\inst{4} \and Sofia {Vazquez Alferez}\inst{2}}
\authorrunning{Bachmann et al.}
\institute{University of Passau, 94032 Passau, Germany 
\and
Department of Information and Computing Sciences, Utrecht University, Princetonplein 5, 3584 CC Utrecht, The Netherlands 
\and
Faculty of Discrete Mathematics and Optimization, FernUniversität in Hagen, Universitätsstraße 1, 58084 Hagen, Germany
\and
TUM School of Computation, Technical University of Munich, Bildungscampus 2, 74076 Heilbronn, Germany
}
\maketitle              %

\begin{abstract}
We show that the \CoinDraw problem is \ER-complete.
In the \CoinDraw problem we are given an arrangement $\A$ of pseudocircles and we ask if there is an equivalent arrangement of unit circles $\C$.
This extends a recent result on the stretchability of pseudocircles of arbitrary radius.
It places the \CoinDraw problem in the same complexity class as other problems about stretchable arrangements of lines [Shor, 1991] and circles [Felsner \& Scheucher, 2020] as well as the recognition of (unit) disk intersection graphs [McDiarmid \& M\"uller, 2013] and unit disk contact graphs [Lubiw \& Schaefer, 2025].
\end{abstract}


\section{Introduction}
Say you are at an open problem workshop in a wonderful historic castle in the Netherlands.
Your toddler comes with you and she wants to be entertained.
So you give her a large coin and ask her to draw some circles using the coin. (Most of the circles will intersect.)
Quickly she draws many circles on a piece of paper meant for a flip chart.
She uses different colors and wonders about all the patterns that she can draw with the one coin.
On  the next day, she comes with a drawing of some random toddler circles (far from perfect circles) and wonders if she could have drawn the same pattern with a coin.
Thus we ask: \emph{how difficult is it to determine whether a pattern of loosely-drawn `circles' can be made using only a single coin?}

\paragraph*{Definitions.}
We formalize this research question with the following definitions.

\begin{figure}
    \centering
    \includegraphics{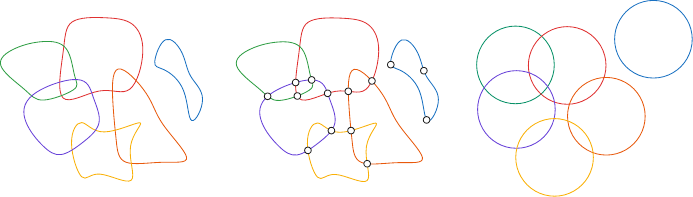}
    \caption{Left: a pseudocircle arrangement \A. Middle: The planar graph $G(\A)$ induced by the pseudocircles. Right: An equivalent arrangement $A$ of unit circles. }
    \label{fig:definition}
\end{figure}

An \textit{arrangement of pseudocircles} \A (see also Figure~\ref{fig:definition})
is a collection of  $n$ simple closed curves in the plane such that any two of them are either disjoint or cross in exactly two points. 
In the following, we use $n$ to denote the number of input pseudocircles.
Each such arrangement induces a planar (multi)graph  
$G = G(\A)$:
Introduce one vertex per crossing and an edge for each curve segment between two consecutive crossings.
For the purpose of this paper, we assume that the arrangement of the pseudocircles is given by a combinatorial embedding of $G$.
That is, next to the abstract graph $G$ itself, we are given for each vertex the order in which the edges are drawn around the vertex.
Furthermore, we label each edge with $1,\ldots, n$ indicating the pseudocircle that it belongs to.
A pseudocircle with no neighbors is modeled as a cycle on three vertices (See Figure~\ref{fig:definition}-Middle).

Two arrangements are \textit{equivalent} if some homeomorphism of the plane takes one to the other.
Equivalently, two arrangements are equivalent if their planar graphs have a combinatorially equivalent embedding.
Linhart–Ortner showed that the two definitions of equivalence coincide~\cite{linhart-ortner-04}.

An arrangement \A of pseudocircles is \emph{penny-realizable} if there is an arrangement $A$ of unit circles that is equivalent to \A. 
The \CoinDraw problem (re-stated immediately below for easy reference) is the problem of deciding whether a given pseudocircle arrangement is penny-realizable.
Note that, by scaling, this is equivalent to using circles with the same diameter.
\begin{problemd}
    \problemtitle{\CoinDraw}
    \probleminput{An arrangement $\A$ of pseudocircles.}
    \problemquestion{Is there an arrangement $A$ of unit circles that is equivalent to $\A$?
    }
\end{problemd}

\paragraph*{Results.}
    We completely settle the computational complexity of \CoinDraw by proving the following theorem.
    \begin{restatable}{theorem}{ercomplete}
        \label{thm:ercomplete}
        \CoinDraw is \ER-complete.
    \end{restatable}

    \noindent The \ER-membership proof is completely standard in the area.
    We regard the \ER-hardness as the novel contribution.

    \begin{figure}
        \centering
        \includegraphics{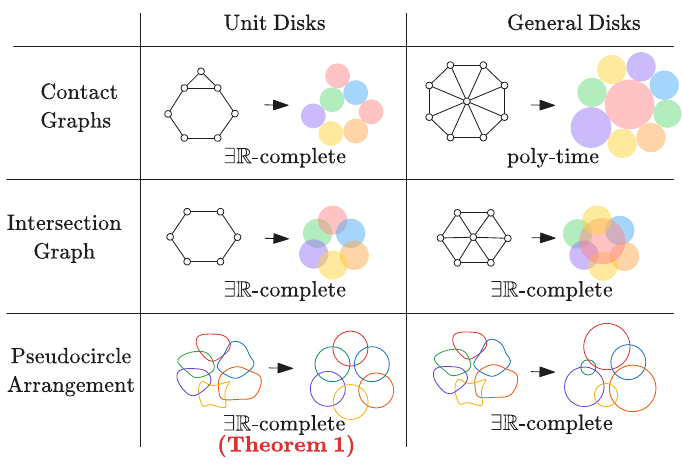}
        \caption{An overview of the different 
         computational problems on representations of drawing with circles from the literature.}
        \label{fig:literature-gap}
    \end{figure}
    
    \paragraph*{Related work.}
    Our result answers one of the many questions one can ask about representations of drawings with circles (see Figure \ref{fig:literature-gap} for a succinct overview of the many variations).
    Many earlier results concern recognition of contact graphs, intersection graphs, and pseudocircle arrangements of (unit) circles.
    We emphasize that none of the problems are a more general case of another nor do any of the proofs carry over from one setting to another setting.
    
    The first major result is Koebe's Kissing Lemma~\cite{koebe-36} --- also known as the Circle Packing Theorem --- from 1936, which asserts that every planar graph is the contact graph of disks.
    Since testing for planarity can be done in linear time (see, for example, Boyer and Myrvold~\cite{Boyer_Myrvold_2004}), recognizing whether a graph can be represented by an arrangement of touching disks is linear time solvable.
    
    However, slight modifications to the problem make it much more difficult.
    Firstly, we can ask whether a graph can be represented by an arrangement of touching \emph{unit} disks, which is also known as recognizing \textit{penny graphs}. 
    This problem is known to be \ER-complete by Lubiw and Schaefer~\cite{lubiw-schaefer-25}.

    Another variation asks whether a graph can be represented by an arrangement of \textit{intersecting} rather than touching (unit) disks.
    In this case, the disks should overlap if and only if their respective vertices are incident to the same edge.
    This variant was proven to be \ER-complete for unit disks by Kang and M\"uller in 2011~\cite{kang-muller-12} and by McDiarmid and M\"uller in 2013 for general disks~\cite{mcdiarmid-muller-13}.

    Finally, it is natural to consider the case where we are not given a graph but an arrangement of pseudocircles.
    Here, our task is to find an equivalent arrangement of overlapping (unit) disks.
    (Note that with our definition of pseudocircles, there can never be an equivalent arrangement of touching (unit) disks.)
    The version for disks with arbitrary radii was proven to be \ER-complete by Kang and M\"uller in 2014~\cite{kang-muller-14}.
    Furthermore, it remains \ER-complete when restricted to great pseudocircles~\cite{felsner-scheucher-20}.
    Although it is also natural to ask whether the question remains \ER-complete when restricted to unit circles, we are not aware of any work that answers this.
    In this paper, we close this gap and present a concise proof of
    the complexity of recognizing whether an arrangement of pseudocircles has an equivalent arrangement of unit circles.

\paragraph*{Existential Theory of the Reals in a Nutshell.}
Many natural problems in geometry are known to be \NP-hard but not known to be in \NP. 
Often, solutions to such problems do exist but they require irrational coordinates, which rules out obvious candidates for an \NP certificate. 
In the complexity class, \emph{existential theory of the reals}, abbreviated $\ER$,
one relaxes the requirement that certificates have polynomially bounded binary representations to admit certificates containing real numbers.
Additionally, for the verifier to handle such real-valued witnesses, we must adjust our computational model: the verifiers work in the {RealRAM model}, in which the WordRAM model is extended with real registers that can be used to store and manipulate real numbers.
In other words, the complexity class \ER captures problems that have polynomial time verifiers but real-valued witnesses \cite{erickson-etal-22}.
Note that instances of problems in \ER all have binary encodings, so reductions in \ER are the standard polynomial time reductions.
It is known that $\NP\subseteq\ER\subseteq\PSPACE$~\cite{canny-88}.

    What makes the complexity class \ER important are its numerous complete problems from various areas of theoretical computer science.
    The compendium from Schaefer, Cardinal and Miltzow lists about 150 of them~\cite{schaefer-cardinal-miltzow-26}.
    Some important examples are training of neural networks~\cite{abrahamsen-kleist-miltzow-21,bertschinger-etal-23}, geometric packing~\cite{abrahamsen-miltzow-seiferth-24}, the art gallery problem~\cite{abrahamsen-adamaszek-miltzow-22} and the recognition of different geometric intersection graphs~\cite{kratochvil-matousek-94,cardinal-etal-18}.

\begin{figure}
        \centering
            \includegraphics{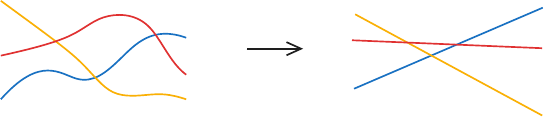}
        \caption{The \stretchability problem can be thought of as a set of strings that are loosely on a table. Stretch each of them to make them straight without changing their intersection pattern.}
        \label{fig:stretchability}
    \end{figure}

    Many of the problems are shown to be \ER-complete by a reduction from \stretchability.
    This makes \stretchability one the most significant \ER-complete problems.
    In \stretchability we are given an arrangement~\L of simple open curves such that any two curves intersect exactly once and no more than two lines intersect in a point.
    We are asked if there is an arrangement of straight lines with the same combinatorics.
    The combinatorics are defined as the order of intersections of each line with respect to each other line, see \cref{fig:stretchability} for an illustration.
    We will use this problem in our reductions as well.

\section{\ER-Membership}
We show \ER-membership using the machine model from~\cite{erickson-etal-22}.
That is, \CoinDraw is in \ER if there exists a verification algorithm \textit{Alg} running in polynomial time in the RealRAM, such that for every binary encoded instance $I$ of \CoinDraw it holds that $I$ is a yes-instance of \CoinDraw if and only if there exists a witness $w\in \R^*$ such that $\textit{Alg}(I,w)=1$.

\label{sec:Membership}
\begin{lemma}\label{thm:membership}
The \CoinDraw problem is in \ER.
\end{lemma}
\begin{proof}
    Let~\A be the input arrangement of $n$ pseudocircles with planar graph $G(\A)$.
    The witness is the set $\{(x_i,y_i)\in \R^2\;|\;i\in\{1,\ldots,n\}\}$ representing the centers of the arrangement \U of unit circles.
    Given this set of points the verification algorithm, working in the RealRAM model, computes the intersections of every pair of circles $(i,j)$.
    The algorithm then uses the circles and the intersections to compute the induced planar graph $G(\U)$ of the \coinArrangement{} along with its combinatorial embedding.
    Then, it checks that $G(\U)$ is combinatorially equivalent to $G(\A)$. 
\end{proof}

\section{\ER-Hardness}
\label{sec:hardness}

In this section, we prove the following lemma:

\begin{lemma}
    \label{lem:erhard}
    The \CoinDraw problem is \ER-hard.
\end{lemma}

To prove this, we reduce from the \stretchability problem:

\begin{problemd}
    \problemtitle{\stretchability}
    \probleminput{An arrangement $\L$ of simple open curves such that any two curves intersect exactly once and no more than two lines intersect at a single point.}
    \problemquestion{Is there an arrangement of straight lines that is equivalent to $\L$?
    }
\end{problemd}
Throughout the paper we assume that the input arrangement $\mathcal{L}$ is given as a \emph{wiring diagram}. 
A wiring diagram consists of $n$ piece-wise linear $x$-monotone curves, called \emph{wires}, drawn in a horizontal strip and running from its left edge to its right edge, such that any two wires cross exactly once and no three wires meet in a point. 
Each wire occupies one of the $n$ horizontal levels except in a small neighborhood of a crossing, where the two wires involved exchange levels (see \cref{fig:construction}-Left for an example).
Wiring diagrams and simple arrangements of pseudolines are equivalent representations\cite{goodman1980,bjorner1999}, %
 and \stretchability is $\exists\mathbb{R}$-complete already for inputs given as wiring diagrams~\cite{mnev-88,shor-91}.
\subsection{Intuition}
\label{sec:intuition}
\begin{figure}
    \centering
    \includegraphics[width=0.9\linewidth]{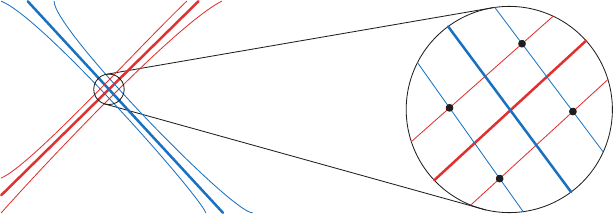}
    \caption{Pseudocircle simulation (thin lines) of two crossing straight lines (thick), alongside zoomed-in views highlighting their four crossing points (indicated by black vertices). }
    \label{fig:sandwichcrossing}
\end{figure}

We begin by giving some intuition about the reduction before we formally describe it. 
We need to encode a given pseudoline arrangement with pseudocircles.
The idea is that we can ``simulate'' a given pseudoline by a pair of disjoint pseudocircles that sandwich the pseudoline between them (one `above' the pseudoline and the other `below') as illustrated in \cref{fig:construction} in \cref{sec:construction} -- this encoding is similar to the one used by \cite{felsner-scheucher-20}.
In such a simulation, each crossing of two pseudolines is replaced by four crossings of closed curves, see \cref{fig:sandwichcrossing}. 
Thus, we can recover all combinatorial information about the crossings of pseudolines from corresponding crossings of pseudocircles.
(Notice that because all pseudolines cross each other, a pseudocircle will intersect all other pseudocircles except for its disjoint pair.)
Of course, two pseudocircles that are not disjoint cross exactly twice, so for each pair of non-disjoint pseudocircles we must specify a second crossing.
This must be done in a way that guarantees two things: 
(a) that given a penny arrangement realization of a pseudocircle encoding of \stretchability, 
we can still distinguish which crossings contain the combinatorial information about the pseudoline arrangement; and
(b) that whenever the encoded \stretchability instance is a yes-instance, the second crossings do not introduce unrealizable combinatorics for a penny arrangement.

We begin by addressing (a). %
With sufficiently large radii, we can find a bounded convex region, $\bbox$, of the plane where (i) the circular segments of disjoint pairs behave like straight lines combinatorially and (ii) $\bbox$ contains all the crossings encoding pseudoline crossings (and no other crossings).
Indeed, a solution of \stretchability has no crossings at infinity so the corresponding crossings of circles can be contained in some finite region of the plane.
Then, by increasing the radii of our circles, we can make sure that the second crossings are as ``far away'' as we want from the crossings encoding \stretchability.

We work towards (b) next.
If we take the bisector of each disjoint pair of circles, we obtain a straight-line solution to \stretchability inside $\bbox$; and if we zoom-out, each circle of a disjoint pair appears tangent to this bisector.
Zoom out further until the area of $\bbox$ is negligible. Then, all circles appear to intersect on one point (the collapsed $\bbox$) and they rotate counterclockwise around $\bbox$ according to the slope of the bisectors: from smallest slope to largest slope, starting with the circle `above' the bisector with smallest slope and ending with the circle `below' the bisector with the largest slope.
We can tell what the order of these slopes is without knowing the solution to a \stretchability instance!
Indeed, this is just the top to bottom order of the pseudolines at the left edge of the wiring diagram.
So we can use this order to specify the second crossings in our construction.  
In what follows, we call crossings inside $\bbox$ \emph{\intersectionsA} and crossings outside $\bbox$ \emph{\intersectionsB}.

\subsection{Construction of the Reduction}
\label{sec:construction}
\begin{figure}
    \centering
    \includegraphics[width=1.0\linewidth]{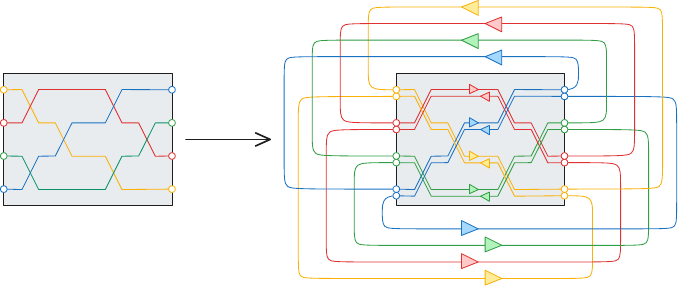}
    \caption{Left: An instance of \stretchability $\mathcal{L}$ with four pseudolines given as a wiring diagram. Right: An instance $\mathcal{A}$ of \CoinDraw that can be used to solve whether $\mathcal{L}$ is stretchable. Note that $\mathcal{A}$ has eight pseudocircles. The arrows show the counterclockwise orientation of each pseudocircle. }
    \label{fig:construction}
\end{figure}

Given any instance $\mathcal{L}$ of \stretchability, we construct an instance $\mathcal{A}$ of \CoinDraw such that there exists a penny arrangement $A$ equivalent to $\mathcal{A}$ if and only if there exists a line arrangement $L$ equivalent to $\mathcal{L}$.
This construction is 
depicted
in Figure \ref{fig:construction}.

We index the pseudolines $\psline{1}, \ldots, \psline{n}$ of $\mathcal{L}$ by the levels they occupy at the left edge of the diagram, from top to bottom. 
Since every pair of wires crosses exactly once, the order at the right edge is the reverse, $\psline{n},\dots,\psline{1}$. In a straight-line realization the
left-to-right sweep direction corresponds to reading the lines from
$x=-\infty$ to $x=\infty$, so this indexing lists the lines by increasing slope.
In
particular, for $i<j$ the line $\psline{i}$ lies above $\psline{j}$ to the left of their intersection and below it to the right. We refer to this indexing as the \emph{slope order} of $\mathcal{L}$.

For each pseudoline~$\psline{i}$ in $\mathcal{L}$, we add two mutually disjoint pseudocircles~$\pscircUp{i}$ and $\pscircDown{i}$ to $\mathcal{A}$, with $\pscircUD{i}:=\{ \pscircUp{i}, \pscircDown{i} \}$ for short.
The pseudocircle~$\pscircUp{i}$ lies on top of $\psline{i}$ and is refered to as the \emph{top} copy, whilst $\pscircDown{i}$ lies below $\psline{i}$ and is refered to as the \emph{bottom} copy.
More formally, for every point $(x,y)\in \ell_i$ the following are true: if $(x,y')\in \pscircUp{i}$ then $y'>y$; similarly, if $(x,y'')\in \pscircDown{i}$ then $y>y''$.
In Figure \ref{fig:construction}, the top and bottom copies are drawn in the same color as their corresponding pseudoline.

Whilst the pseudocircles $\pscircUD{i}$ are mutually disjoint they will both intersect every other pseudocircle twice.
The rest of this section specifies the order in which these intersections occur.
For this, we orient every pseudocircle of $\mathcal{A}$ counterclockwise, so all cyclic orders specified below are read in this direction.
Half the intersections of the pseudocircles are specified according to the intersections of $\L$ --- which we call the \emph{\intersectionsA}.
The other half of the intersections ---which we called \emph{\intersectionsB}--- are specified using \emph{only} the indices of the pseudocircles; which carry information about the slope order of $\L$.
Moreover, for any pseudocircle we let all crossings belonging to the \intersectionsA precede, in the cyclic order, any crossing belonging to the \intersectionsB.

First, we specify the \intersectionsA, which encode the intersections of $\L$.
For reference, see the intersections inside the gray box of \cref{fig:construction}; noting that due to the counterclockwise orientation of the pseudocircles, the intersections described below appear from left to right for $\pscircUp{i}$ and from right to left for $\pscircDown{i}$.
Each disjoint pair of pseudocircles $\pscircUD{i}$ in $\A$ will follow the same inner intersection pattern but in reverse order. 
Thus, it suffices to describe the inner intersection pattern of $\pscircUp{i}$:
If $\psline{i}$ intersects~$\psline{j}$ before~$\psline{k}$, then~$\pscircUp{i}$ will intersect~$\pscircUD{j}$ before~$\pscircUD{k}$.
However, this still leaves open whether $\pscircUp{i}$ intersects $\pscircUp{j}$ or $\pscircDown{j}$ first.
To resolve this, we use the slope order of~$\L$.
If $i < j$, then $\pscircUp{i}$ is above $\pscircUD{j}$ before the intersection.
So, it must intersect $\pscircUp{j}$ first, as that is the top copy.
(We emphasize that the intersection pattern of $\pscircDown{i}$ is the reverse of this.)
This concludes the specification of the first half of the intersections of $\A$. 
Indeed, we have described one intersection for each pair of non-disjoint pseudocircles.

The \intersectionsB are specified with the following rule:

Consider a pseudocircle $\pscircUp{i} \in \mathcal{A}$. %
Then, after the \intersectionsA, the order in which $\pscircUp{i}$ intersects the other pseudocircles is the following:
\[ \pscircDown{{i+1}}, \pscircDown{{i+2}}, \dots, \pscircDown{{n}},
  \pscircUp{{1}}, \pscircUp{{2}}, \dots, \pscircUp{{i-1}},
  \pscircUp{{i+1}}, \pscircUp{{i+2}}, \dots, \pscircUp{{n}},
  \pscircDown{{1}}, \pscircDown{{2}}, \dots, \pscircDown{{i-1}}. \]
For its counterpart, $\pscircDown{i}$, we get the following order:
\[ \pscircUp{{i+1}}, \pscircUp{{i+2}}, \dots, \pscircUp{{n}}, \pscircDown{{1}}, \pscircDown{{2}}, \dots, \pscircDown{{i-1}}, \pscircDown{{i+1}}, \pscircDown{{i+2}}, \dots, \pscircDown{{n}}, \pscircUp{{1}}, \pscircUp{{2}}, \dots, \pscircUp{{i-1}}. \]
Each list  has $2n-2$ entries, one for every pseudocircle other than $\pscircUD{i}$, as required.

Everything we have described for the construction is a function of the wiring diagram alone, so $\mathcal{A}$ is well defined and computable in time polynomial in $n$.

\subsection{Correctness of the Reduction}\label{subsec:correctness}

We now argue that the construction from Section \ref{sec:construction} gives a correct reduction.
In other words, $\mathcal{L}$ is a yes-instance for \stretchability if and only if $\mathcal{A}$ (obtained via the reduction) is a yes-instance of \CoinDraw.

\subsubsection{\stretchability $\implies$ \CoinDraw.}
\label{subsec:stretchtocoin}
Let $\mathcal{L}$ be a yes-instance of \stretchability and let $L$ be a solution, i.e. an equivalent arrangement of lines.
We show that the arrangement of pseudocircles $\A$ outputted by our construction from \cref{sec:construction} has an equivalent penny arrangement $A$.

\paragraph{Constructing $A$:} The \intersectionsA of $\mathcal{A}$ are based on $\mathcal{L}$.
Now, we use $L$ to construct a penny arrangement $A$ equivalent to $\mathcal{A}$.
Let $\stline{i} \in L$ be the line that corresponds to the pseudoline $\psline{i} \in \L$.
For each line $\stline{i} \in L$, we add two circles $\stcircUp{i}$ and $\stcircDown{i}$ to $A$, such that $\stline{i}$ is the bisector of $\stcircUp{i}$ and $\stcircDown{i}$.
The exact centers of these circles do not matter so long as: 
(a) some circular arcs of $\stcircUp{i}$ and $\stcircDown{i}$ go through $\bbox$ and, within $\bbox$, have distance at most $\delta$ (defined below) of $\stline{i}$; and
(b) the radii of the circles are large enough such that increasing them further does not change the combinatorics of the penny arrangement.

As $\stcircDown{i}$ and $\stcircUp{i}$ do not intersect, their bisector $\stline{i}$ will not intersect either circle.
Furthermore, this also implies they lie on opposite sides of $\stline{i}$.
We assume that $\stcircUp{i}$ lies `above' $\stline{i}$ and $\stcircDown{i}$ lies `below' $\stline{i}$.
We use $\overline{\underline{c_i}}$ to denote the pair $\{ \overline{c_i}, \underline{c_i} \}$.

\begin{figure}
    \centering
    \includegraphics[width=0.7\textwidth]{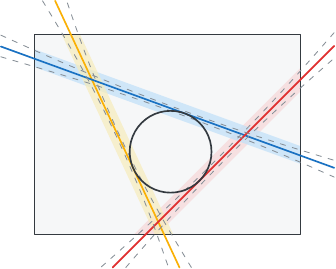}
    \caption{An example of three lines $\stline{i}$ (red), $\stline{j}$ (blue), $\stline{k}$ (yellow) and the inscribed circle $Q_{ijk}$ (black). 
    The circular segments $\circsegUD{i}:=\{\overline{x_i},\underline{x_i}\}$ must be within the red shaded area, $\overline{\underline{x_j}}$ inside the blue, and $\overline{\underline{x_k}}$ inside the yellow --- these circular segments are indicated with dashed lines.
    The shaded regions have width $< 2\cdot \delta$.
    }
    \label{fig:rectangle}
\end{figure}

\noindent To show the equivalence between $\A$ and $A$ we distinguish between two ways in which pseudocircles intersect: \intersectionsA and \intersectionsB.
We recall that \bbox is the convex region in which all (and only) \intersectionsA occur; and that $B$ covers a finite region of the plane because $L$ does not contain any intersections at infinity.
This gives us two cases.
\paragraph{The inner part:}
We prove that the \intersectionsA must be exactly the intersections specified in the construction (\cref{sec:construction}).

Firstly, consider the way $\stcircUp{i}$ intersects $\stcircUD{j}$.
We can assume $i < j$ as the case $i>j$ is analogous but yields the exact reverse order of intersections.
We need to show that $\stcircUp{i}$ intersects $\stcircUp{j}$ before it intersects $\stcircDown{j}$.
Towards a contradiction, assume otherwise.
As $\stcircUp{i}$ is above $\stcircUD{j}$ before the intersections, $\stcircDown{j}$ must be above $\stcircUp{j}$ at its intersection.
This also implies that $\stcircDown{j}$ is above $\stline{j}$.
This contradicts the definition of $\stcircDown{j}$ (as  the circle below $\stline{j}$).
The argument for how $\stcircDown{i}$ intersects $\stcircUD{j}$ with $i<j$ is analogous but $\stcircDown{i}$ intersects $\stcircDown{j}$ before $\stcircUp{j}$.

Secondly, if $\stline{i}$ intersects $\stline{j}$ before $\stline{k}$, we show that $\stcircUp{i}$ must intersect $\stcircUD{j}$ before $\stcircUD{k}$.
Without loss of generality, assume $j<i<k$, so by the previous paragraph, $\stcircUp{i}$ intersects $\stcircDown{j}$ before $\stcircUp{j}$, and also $\stcircUp{k}$ before $\stcircDown{k}$.
We prove that $\stcircUp{i}$ intersects $\stcircUp{j}$ before $\stcircUp{k}$, implying it intersects $\stcircUD{j}$ before $\stcircUD{k}$.

Consider any triplet of lines in $\stline{i}, \stline{j}, \stline{k} \in L$.
Let $\delta_{ijk}$ radius of the inscribed circle $Q_{ijk}$, 
and define $\delta = \min_{i,j,k} \delta_{ijk}$\label{delta}.
For a circle $c_i$ let $x_i = c_i \cap \bbox$ be the circular arc segment of $c_i$'s boundary contained in $B$.
The circles $\stcircUD{i}$ have radius large enough such that
the distance between any point $p\in\overline{x_i}$ (respectively, any point $p'\in \underline{x}_i$) and $\stline{i}$ is strictly smaller than $\delta$ (this is the $\delta$ that was mentioned in \cref{sec:intuition}). See \cref{fig:rectangle} for reference.

If $\stcircUp{i}$ intersects $\stcircUp{k}$ before $\stcircUp{j}$, 
then both intersections must at distance $< \delta$ of $\stline{i}$, $\stline{j}$ and $\stline{k}$.
Indeed, the intersection $\stcircUp{i}\cap\stcircUp{j}$ is, by the previous paragraph, within distance $\delta$ of both $\stline{i}$ and $\stline{j}$.
Furthermore, since $\stcircUp{i}$ intersects $\stcircUp{k}$ before $\stcircUp{j}$ we have that $\stcircUp{i}\cap\stcircUp{j}$ is further towards to $\stline{k}$ than the intersection $\stcircUp{i}\cap\stcircUp{k}$ --- i.e. is at distance $< \delta$ of $\stline{k}$.
However, if any point lies at distance $< \delta$ of all three lines, then the center of the circle $Q_{ijk}$ also lies within distance $< \delta$ of all three lines.
This implies that $\delta_{i, j, k} < \delta$, which is a contradiction, as $\delta$ is defined as the minimum radius across all inscribed circles.
Thus, the order of intersections is maintained.

\paragraph{The outer part:} 
\begin{figure}
    \centering
    \includegraphics[]{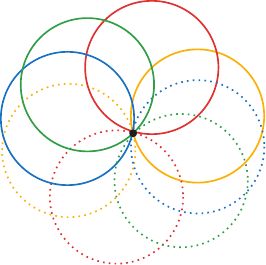}
    \caption{An example of four pairs of top and bottom circles which intersect in $\bbox$. The bottom circles are drawn with dotted lines. Note that the instance is homeomorphic to the one from Figure \ref{fig:construction}.}
    \label{fig:Disks}
\end{figure}
We have shown that a yes-instance for \stretchability implies that there exists an arrangement of circles $\C$ such that the \intersectionsA inside \bbox are valid;
now we need to argue that the outer part does not invalidate the instance.

Let $C$ be a circle of radius $r'$ containing $B$.
Since $B$ is a finite region of the plane, we treat $r'$ as a constant.
As the radius $r$ of the circles in $A$ becomes larger and larger, the distance between the \intersectionsA becomes negligible.
(Note that we can do this without affecting \intersectionsA;
the arcs within \bbox will simply be straighter and thus stay within distance $\delta$ of the lines in $L$.)
Specifically, $\lim_{r \to \infty} \frac{r'}{r} = 0$.
Then, since all \intersectionsA are within $B$, for sufficiently large $r$, the region $B$ can be treated as a point.
Thus, for simplicity, we pretend that all the circles in $A$ intersect in a singular point, \bbox.

As the circles in $\C$ all have the same radius, their centers will lie on a circle of radius $r$ around $\bbox$ (see Figure \ref{fig:Disks} for intuition).
Furthermore, using the logic from the construction, the lines in $L$ will be tangents to the circles.
Using these tangents, we deduce the order of the centers of circles in $\C$ in counterclockwise order around \bbox.
This yields a highly structured instance, which we use to determine the order of their intersections.

Without loss of generality, we choose any of the `top' disks $\stcircUp{i}$.
By the construction, we see it is supposed to intersect the other circles in the following order:
\[ \stcircDown{{i+1}}, \stcircDown{{i+2}}, \dots, \stcircDown{{n}},
  \stcircUp{{1}}, \stcircUp{{2}}, \dots, \stcircUp{{i-1}},
  \stcircUp{{i+1}}, \stcircUp{{i+2}}, \dots, \stcircUp{{n}},
  \stcircDown{{1}}, \stcircDown{{2}}, \dots, \stcircDown{{i-1}}. \]

If we consider a `bottom' circle $\stcircDown{i}$ instead, we get the following order:

\[ \stcircUp{{i+1}}, \stcircUp{{i+2}}, \dots, \stcircUp{{n}}, \stcircDown{{1}}, \stcircDown{{2}}, \dots, \stcircDown{{i-1}}, \stcircDown{{i+1}}, \stcircDown{{i+2}}, \dots, \stcircDown{{n}}, \stcircUp{{1}}, \stcircUp{{2}}, \dots, \stcircUp{{i-1}}. \]

Observe that both orders corresponds to exactly the counterclockwise order of the tangents through \bbox starting at (but skipping) $\stcircUD{i}$.
This is also equivalent to the counterclockwise order of the centers of the circles around $B$.
For this reason, we argue that this order is correct, and ignore whether circles are `top' or `bottom' ones.

\begin{figure}[h!]
    \centering
    \includegraphics[]{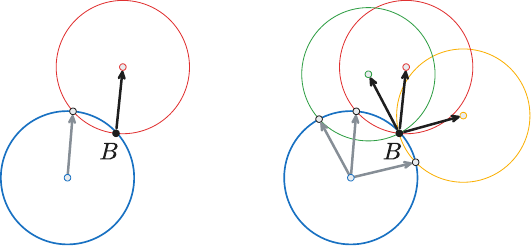}
    \caption{Left: A pair of disks with two intersections.
    Note how the vector between the common intersection $B$ to the blue center is the same as the vector from the red center to the `outer' intersection.
    Right: The intersection order around the blue disk matches the rotation order of the centers around $B$.}
    \label{fig:vectors}
\end{figure}

Let circle $\stcirc{i}$ intersect some circle $x$ and let $v(a, b)$ denote the vector from $a$ to $b$.
The intersection point outside of \bbox between $c_i$ and $x$ is exactly $\textsc{Center}(\stcirc{i}) + \allowbreak v(\bbox, \allowbreak \textsc{Center}(x))$, as shown in Figure \ref{fig:vectors}.
As $v(\bbox, \textsc{Center}(x))$ is always of length $r$, it only varies in its rotation.
So, as $v(\bbox, \textsc{Center}(x))$ rotates around \bbox, the intersections rotate in the same order around $\textsc{Center}(\stcirc{i})$.

\subsubsection{\CoinDraw $\implies$ \stretchability.}
\label{subsec:cointostretch}
Let $\C$ be a \coinArrangement that is a solution to instance $\A$ of \CoinDraw.
We now construct a line arrangement $L$ that is a solution for instance \L of \stretchability.

\begin{figure}[h!]
    \centering
    \includegraphics[]{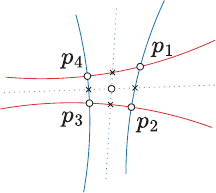}
    \caption{An example of a kite between two pairs of top and bottom circles (drawn in blue and red) along with their bisectors.}
    \label{fig:kite}
\end{figure}

For each $\stcircUp{i}\in\C$ there exists exactly one other circle $\stcircDown{i}$ in the arrangement such that $\stcircUp{i}\cap \stcircDown{i}=\emptyset$.
For each such pair of circles 
$\stcircUD{i}$ we construct their bisector $\stline{i}$.
We claim that the set $L=\{\stline{i}|\;i\in [n]\}$ of all such $\stline{i}$ is the desired solution of instance $\L$ of \stretchability.

Take any two pairs of top and bottom circles $\stcircUD{i},\stcircUD{j}$ , and let $b_i, b_j$ be the bisectors of each pair respectively. 
By definition, each of the four pairs of intersecting circles from the $\stcircUD{i} \cup \stcircUD{j}$ has two intersection points.
Let $p_1$ be the closest $\stcircUp{i} \cap \stcircUp{j}$-intersection to the crossing $b_i \cap b_j$.
Similarly, let $p_2\in\stcircUp{i} \cap \stcircDown{j}, p_3\in\stcircDown{i} \cap \stcircDown{j}, p_4\in\stcircDown{i} \cap \stcircUp{j}$ be the intersections closest to the $b_i \cap b_j$ crossing for the corresponding pairs of intersecting circles.
(These will be exactly the \intersectionsA of $\stcircUD{i} \cup \stcircUD{j}$.)
We call the region enclosed by the circular arcs between $p_1\ldots p_4$ a \textit{kite} (see Figure \ref{fig:kite}), and for a pair $\stcircUD{i}, \stcircUD{j}$ we denote it $K_{ij}$.

Note that by construction, any circle $c_i$ in $\stcircUD{i}$ must intersect $\stcircUD{j}$ consecutively.
Therefore, for all $\stcirc{k}$ with $i\neq k\neq j$ it holds that $K_{ij}\cap c_k=\emptyset$.
Indeed, $K_{ij}\cap c_k\neq\emptyset$ implies that $c_k$ intersects at least one edges of $K_{ij}$.
Without loss of generality, assume it intersects $p_1p_2 \in \stcircUp{i}$.
This contradicts that the intersections $p_1, p_2$ are consecutive along $\stcircUp{i}$.
Thus, any bisector $\stline{i}$, oriented from left to right, intersects the circles in the same order as $\stcircUp{i}$ (equiv. in reverse order as $\stcircDown{i}$).
Furthermore, it intersects the bisector of $\stcircUD{j}$ in between its intersections with those two circles.
Thus, the intersection order of bisector $\stline{i}$ with bisectors $\{\stline{j}|\;i\neq j\}$ is combinatorially the same as the order in which $\stcircUp{i}$ intersects $\stcircUp{j}$ for $i\neq j$.
Thus, the set $\{ \stline{1} \dots \stline{n}\}$ has combinatorially the same intersections as $\C$, which in turn is combinatorially identical to $\A$, which by the reduction is combinatorially identical to $\L$.
Therefore, have constructed a stretched line arrangement of $\L$.

Both arguments of Section~\ref{subsec:correctness} together conclude the proof of Lemma~\ref{lem:erhard}. 

\section{Concluding Remarks}

    The \ER-membership from Section~\ref{sec:Membership} and the hardness result from Section~\ref{sec:hardness} yield our main theorem:
    \ercomplete* 
    
    \ER-completeness is often described as a level of difficulty that excludes combinatorial algorithms from solving the problem exactly.
    Some problems in \ER have instances where any witness has an exponential number of bits in their binary representation (some witnesses can even be shown to be necessarily irrational).
    This means that unless $\NP = \ER$, there are instances of \CoinDraw where the centers of solution circles do not have short representations --- i.e. the centers of the circles require exponential bit complexity to describe. 
    
    It is often easier to reduce problem A to problem B when A and B share some common objects. 
    Since there are many interesting problems that ask about the complexity of closed curves~\cite{erickson2019} and unit disks~\cite{lubiw-schaefer-25}, we hope \CoinDraw can be an easy starting point for reductions on such problems.
    In spirit, the \CoinDraw problem is very similar to \stretchability; it only exchanges (pseudo)lines for unit (pseudo)circles.
    Since \stretchability is used in many reductions to prove \ER-hardness we hope that \CoinDraw will be similarly manageable.

    Finally, we found the proof of \ER-completeness here to be surprisingly concise, intuitive, and visual.
    This makes the proof well-suited for teaching material:
    One could replace some of the textual details of why the circles intersect with a few pictures that illustrate the intuition behind it; and leave filling up the details as an exercise.
    \paragraph{Open Problem:}
    We wonder about the complexity of the following problem: given an arrangement of pseudocircles $a_1, \dots, a_n$ and a set of points $p_1, \dots, p_n$ is there an equivalent arrangement of (not necessarily unit) circles $c_1, \dots, c_n$ such that circle $c_i$ represents $a_i$ and has its center at $p_i$?

    If this problem is not \ER-hard, then with Theorem \ref{thm:ercomplete} we can conclude that the \ER-hardness of (Pseudocircle) Circularizability~\cite{schaefer-cardinal-miltzow-26} is solely from determining the centers of the circles instead of the radii.
    (Note that, if we are given a disk intersection graph instead of a pseudocircle arrangement, then the open problem is polynomial-time solvable on the RealRAM.)

\begin{credits}
\subsubsection{\ackname} 
    We are grateful for the organizers of GGWeek 2026 that provided a great working environment in the beautiful Kasteel van Oijen.

\subsubsection{AI Usage.}    
    We acknowledge that we used AI to find the open problem in the literature and further related work.
    Furthermore, AI gave feedback on the write-up. 
    However, all proof ideas are human.

\subsubsection{\discintname}
The authors have no competing interests to declare that are
relevant to the content of this article. 
\end{credits}
\bibliographystyle{splncs04}
 \bibliography{coins2}

\begin{thebibliography}{10}
\providecommand{\url}[1]{\texttt{#1}}
\providecommand{\urlprefix}{URL }
\providecommand{\doi}[1]{https://doi.org/#1}

\bibitem{abrahamsen-adamaszek-miltzow-22}
Abrahamsen, M., Adamaszek, A., Miltzow, T.: {The Art Gallery Problem is $\exists\mathbb{R}$-complete}. J. ACM  \textbf{69}(1),  Art. 4, 70 (2022). \doi{10.1145/3486220}

\bibitem{abrahamsen-kleist-miltzow-21}
Abrahamsen, M., Kleist, L., Miltzow, T.: {Training Neural Networks is {ER}-complete}. In: Ranzato, M., Beygelzimer, A., Dauphin, Y.N., Liang, P., Vaughan, J.W. (eds.) Advances in Neural Information Processing Systems 34: Annual Conference on Neural Information Processing Systems 2021, NeurIPS 2021, December 6-14, 2021, virtual. pp. 18293--18306 (2021), \url{https://proceedings.neurips.cc/paper/2021/hash/9813b270ed0288e7c0388f0fd4ec68f5-Abstract.html}

\bibitem{abrahamsen-miltzow-seiferth-24}
Abrahamsen, M., Miltzow, T., Seiferth, N.: {Framework for $\exists\mathbb{R}$-Completeness of Two-Dimensional Packing Problems}. TheoretiCS  \textbf{3},  Art. 11, 78 (2024). \doi{10.46298/theoretics.24.11}

\bibitem{bertschinger-etal-23}
Bertschinger, D., Hertrich, C., Jungeblut, P., Miltzow, T., Weber, S.: {Training Fully Connected Neural Networks is {\(\exists\)}{R}-Complete}. In: Oh, A., Naumann, T., Globerson, A., Saenko, K., Hardt, M., Levine, S. (eds.) Advances in Neural Information Processing Systems 36: Annual Conference on Neural Information Processing Systems 2023, NeurIPS 2023, New Orleans, LA, USA, December 10 - 16, 2023 (2023), \url{http://papers.nips.cc/paper\_files/paper/2023/hash/71c31ebf577ffdad5f4a74156daad518-Abstract-Conference.html}

\bibitem{bjorner1999}
Bj\"orner, A., Vergnas, M.L., Sturmfels, B., White, N., Ziegler, G.M.: {Oriented Matroids}, Encyclopedia of Mathematics and its Applications, vol.~46. Cambridge University Press, Cambridge, 2 edn. (1999). \doi{10.1017/CBO9780511586507}

\bibitem{Boyer_Myrvold_2004}
Boyer, J.M., Myrvold, W.J.: {On the Cutting Edge: Simplified {$O(n)$} Planarity by Edge Addition}. J. Graph Algorithms Appl.  \textbf{8}(3),  241--273 (2004). \doi{10.7155/jgaa.00091}

\bibitem{canny-88}
Canny, J.: {Some algebraic and geometric computations in {PSPACE}}. In: Proc. 20th Annual ACM Symposium on Theory of Computing (STOC). pp. 460--469. ACM (1988). \doi{10.1145/62212.62257}

\bibitem{cardinal-etal-18}
Cardinal, J., Felsner, S., Miltzow, T., Tompkins, C., Vogtenhuber, B.: {Intersection Graphs of Rays and Grounded Segments}. J. Graph Algorithms Appl.  \textbf{22}(2),  273--295 (2018). \doi{10.7155/jgaa.00470}

\bibitem{erickson2019}
Erickson, J.: {Optimal Curve Straightening is $\exists\mathbb{R}$-Complete}. arXiv:1908.09400 (2019). \doi{10.48550/arXiv.1908.09400}

\bibitem{erickson-etal-22}
Erickson, J., van~der Hoog, I., Miltzow, T.: {Smoothing the Gap Between {NP} and $\exists\mathbb{R}$}. SIAM J. Comput.  \textbf{53}(6),  FOCS20--102--FOCS20--138 (2024). \doi{10.1137/20M1385287}

\bibitem{felsner-scheucher-20}
Felsner, S., Scheucher, M.: {Arrangements of Pseudocircles: On Circularizability}. Discrete Comput. Geom.  \textbf{64}(3),  776--813 (2020). \doi{10.1007/s00454-019-00077-y}

\bibitem{goodman1980}
Goodman, J.E.: {Proof of a conjecture of {Burr}, {Gr\"unbaum}, and {Sloane}}. Discrete Math.  \textbf{32}(1),  27--35 (1980). \doi{10.1016/0012-365X(80)90096-5}

\bibitem{kang-muller-12}
Kang, R.J., M{\"u}ller, T.: {Sphere and Dot Product Representations of Graphs}. Discrete Comput. Geom.  \textbf{47}(3),  548--568 (2012). \doi{10.1007/s00454-012-9394-8}

\bibitem{kang-muller-14}
Kang, R.J., M{\"u}ller, T.: {Arrangements of Pseudocircles and Circles}. Discrete Comput. Geom.  \textbf{51}(4),  896--925 (2014). \doi{10.1007/s00454-014-9583-8}

\bibitem{koebe-36}
Koebe, P.: {Kontaktprobleme der konformen Abbildung}. Ber. Verh. S{\"a}chs. Akad. Wiss. Leipzig, Math.-Phys. Kl.  \textbf{88},  141--164 (1936), no DOI: predates DOI registration and is not in Crossref. Predates Mathematical Reviews; not in MathSciNet.

\bibitem{kratochvil-matousek-94}
Kratochv{\'i}l, J., Matou{\v s}ek, J.: {Intersection Graphs of Segments}. J. Combin. Theory Ser. B  \textbf{62}(2),  289--315 (1994). \doi{10.1006/jctb.1994.1071}

\bibitem{linhart-ortner-04}
Linhart, J., Ortner, R.: {On the Combinatorial Structure of Arrangements of Oriented Pseudocircles}. Electron. J. Combin.  \textbf{11}(1),  Research Paper 30, 13 (2004). \doi{10.37236/1783}

\bibitem{lubiw-schaefer-25}
Lubiw, A., Schaefer, M.: {Recognizing Penny and Marble Graphs Is Hard for Existential Theory of the Reals}. In: 34th International Symposium on Graph Drawing and Network Visualization (GD 2026). LIPIcs. Leibniz Int. Proc. Inform., vol.~396, pp. Paper No. 8, 20. Schloss Dagstuhl. Leibniz-Zent. Inform., Wadern (2026). \doi{10.4230/lipics.gd.2026.8}

\bibitem{mcdiarmid-muller-13}
McDiarmid, C., M{\"u}ller, T.: {Integer realizations of disk and segment graphs}. J. Combin. Theory Ser. B  \textbf{103}(1),  114--143 (2013). \doi{10.1016/j.jctb.2012.09.004}

\bibitem{mnev-88}
Mn{\"e}v, N.E.: {The universality theorems on the classification problem of configuration varieties and convex polytopes varieties}. In: Topology and Geometry --- Rohlin Seminar, Lecture Notes in Math., vol.~1346, pp. 527--543. Springer, Berlin (1988). \doi{10.1007/BFb0082792}

\bibitem{schaefer-cardinal-miltzow-26}
Schaefer, M., Cardinal, J., Miltzow, T.: {The Existential Theory of the Reals as a Complexity Class: A Compendium}. In: Building Bridges III, pp. 167--313. Bolyai Soc. Math. Stud., Springer (2026). \doi{10.1007/978-3-032-10503-5_5}

\bibitem{shor-91}
Shor, P.W.: {Stretchability of pseudolines is {NP}-hard}. In: Applied Geometry and Discrete Mathematics, DIMACS Ser. Discrete Math. Theoret. Comput. Sci., vol.~4, pp. 531--554. Amer. Math. Soc., Providence, RI (1991). \doi{10.1090/dimacs/004/41}

\end{thebibliography}

\end{document}